\documentstyle[epsfig,natbib2,natbibmnfix]{mn}

\newcommand{\be}{\begin{equation}}
\newcommand{\ee}{\end{equation}}

\def\ltsima{$\; \buildrel < \over \sim \;$}
\def\simlt{\lower.5ex\hbox{\ltsima}}
\def\gtsima{$\; \buildrel > \over \sim \;$}
\def\simgt{\lower.5ex\hbox{\gtsima}}

\title[The dark matter halos of dwarf galaxies]{The dark matter halos of 
  dwarf galaxies: a challenge for the $\Lambda$CDM paradigm?}

\author[Ferrero et al.]{
\parbox[t]{\textwidth}{
Ismael Ferrero$^{1,2}$, 
Mario G. Abadi$^{1,2}$, 
Julio F. Navarro$^{3}$, 
Laura V. Sales$^{4}$ and\\ 
Sebasti\'an Gurovich$^{1,2}$
}
\\
\\
$^{1}$Instituto de Astronom{\'i}a Te{\'o}rica y Experimental (IATE), Laprida 922 X5000BGR C\'ordoba, Argentina\\
$^{2}$Observatorio Astron\'omico de C\'ordoba and CONICET, Laprida 854 X5000BGR C\'ordoba, Argentina\\
$^{3}$Department of Physics and Astronomy, University of Victoria, Victoria, BC V8P 5C2, Canada\\
$^{4}$Max Planck Institute for Astrophysics, Karl-Schwarzschild-Strasse 1, 85740 Garching, Germany\\
}
\begin{document}

\date{}
\pubyear{2012}
\maketitle

\begin{abstract}
  The cold dark matter halo mass function is much steeper than the
  galaxy stellar mass function on galactic and subgalactic
  scales. This difference is usually reconciled by assuming that the
  galaxy formation efficiency drops sharply with decreasing halo mass,
  so that virtually no dwarf galaxies form in halos less massive than
  $\sim 10^{10} \, M_\odot$.  In turn, this implies that, at any given
  radius, the dark mass enclosed by a galaxy must exceed a certain
  minimum. We use rotation curves of dwarf galaxies compiled from the
  literature to explore whether their enclosed mass is consistent with
  these constraints.  We find that almost one half of the dwarfs in
  our sample with stellar mass in the range $10^6< M_{\rm
    gal}/M_\odot<10^{7}$ are at odds with this restriction: either
  they live in halos with masses substantially below $10^{10}\,
  M_\odot$ or there is a mechanism capable of reducing the dark mass
  enclosed by some of the faintest dwarfs. Neither possibility is
  easily accommodated within the standard $\Lambda$CDM
  scenario. Extending galaxy formation to halos well below $10^{10} \,
  M_\odot$ would lead  to large numbers of dwarf galaxies in excess of
  current estimates; at the same time, the extremely
  low stellar mass of the systems involved makes it unlikely that
  baryonic effects can reduce their dark matter content. Resolving
  this challenge seems to require new insights into dwarf galaxy
  formation, or  perhaps a radical revision of the prevailing paradigm.
\end{abstract}

\section{Introduction}
\label{SecIntro}

Cosmological N-body simulations and theoretical insight have led to
clear predictions for the mass function of dark matter halos that form
in the current $\Lambda$CDM paradigm
\citep{Press1974,Sheth2001,Jenkins2001,Springel2005a}. On galactic and
subgalactic scales, this halo mass function is much steeper than the
galaxy stellar mass function, suggesting a complex non-linear relation
between the mass of a galaxy and that of its surrounding halo.

The need for such non-linear correspondence was recognized in early
attempts to model hierarchical galaxy formation
\citep[e.g.,][]{White1978}, and has been traditionally thought to
imply that the ``efficiency'' of galaxy formation (i.e., the fraction
of baryons in a halo that gets turned into stars and assembled into a
galaxy) decreases steadily with decreasing halo mass so that
effectively few galaxies, if any, form in halos below a certain
minimum mass. With slight variations, this assumption has been a
cornerstone of semi-analytic galaxy formation models
\citep[e.g.,][]{Kauffmann1993,Cole1994,Somerville1999}, and underpins
most attempts to reconcile the shallow faint end of the galaxy
luminosity function with the steep slope of the halo mass function.

The latest results from the Sloan Digital Sky Survey (SDSS) have
extended the galaxy stellar mass function down to $\sim 10^7\,
M_\odot$ \citep{Baldry2008,Li2009}, and have led to even stricter
constraints on how galaxies populate low mass halos. For example,
using either simple abundance-matching techniques or a full-blown
semianalytic model applied to large cosmological N-body simulations,
\citet{Guo2010,Guo2011} conclude that galaxies with stellar mass,
$M_{\rm gal}$, exceeding $\sim 10^6 \, M_\odot$ must inhabit halos
with virial\footnote{We compute all virial quantities within spheres
  of mean density 200 times the critical density for closure.} mass,
$M_{200}$, typically exceeding $10^{10}\, M_\odot$. 

The steep decline in the efficiency of galaxy formation near this
minimum halo mass also implies that most faint galaxies must be
surrounded by halos spanning a small mass range. In the model of
\citet{Guo2010}, for example, the halo mass of dwarfs in the stellar
mass range $10^6<M_{\rm gal}/M_\odot<10^8$ differ by
less than a factor of $\sim 5$.  These results provide readily
testable predictions that have elicited some tension in the
theoretical interpretation of available data on dwarfs.

For example, since dwarf galaxies tend to be dark matter dominated
then having similar halos means that their rotation velocity should
approach a characteristic value of order $30$ km/s, the virial
velocity of a halo of mass $\sim 10^{10} \, M_\odot$. This
would result in a large number of dwarfs with that characteristic
velocity or, equivalently, in a very steep dependence of the number of
galaxies on rotation speed at the faint end, an effect that can be
searched for in blind HI surveys such as {\sc hipass}
\citep{Barnes2001} or {\sc alfalfa} \citep{Giovanelli2005}. The
``velocity width function'' of galaxies reported by such surveys,
however, is much shallower than expected in the scenario outlined
above, and shows no sign of a characteristic velocity
\citep{Zwaan2010,Papastergis2011}.

A related problem has recently been highlighted by
\citet{Boylan-Kolchin2011} in the context of Milky-Way
satellites. These authors note that the kinematics and structure of
dwarf spheroidal (dSph) galaxies suggest that they inhabit halos with
circular velocities well below $30$ km/s. This represents a challenge
not only because it would mean that galaxies {\it do} form in low-mass
halos, but also because, according to the latest N-body simulations,
Milky Way-sized halos should host several massive subhalos which,
apparently, have failed to form visible satellites \citep[see
also][]{Parry2011,Boylan-Kolchin2012,diCintio2011,Vera-Ciro2012}.

In principle, these difficulties can be explained away with plausible
arguments. For example, the dwarf spheroidal companions of the Milky
Way have likely been orbiting in the tidal field of the Galaxy for
several Gyrs. Their dark matter content could therefore have been
affected by tidal stripping, thus hindering the interpretation of
their inferred halo masses. One should also keep in mind that the
apparent conflict concerns a small number of objects, and is therefore
subject to substantial uncertainty. Good mass estimates are only
available for nine Milky Way dSphs, and the theoretical comparison is
based on just seven $\Lambda$CDM halo realizations, six from the
Aquarius Project \citep{Springel2008} plus the Via Lactea simulation
\citep{Diemand2007}. The possibility that the Milky Way is simply an
outlier either in halo mass or in subhalo content thus remains
\citep[][]{Wang2012}.

Further, as discussed by \citet{Papastergis2011}, the velocity-width
function discrepancy could be explained if the gas rotation velocity
systematically underestimates the circular velocity of the surrounding
dark halo. This may occur if the size of the galaxy is small relative
to the radius where a halo reaches its characteristic
velocity \citep{Stoehr2002}. Indeed, the circular velocity of cold dark matter halos
rises gradually with radius \citep[][hereafter
NFW]{Navarro1996,Navarro1997}: a $10^{10} \, M_\odot$ halo typically
reaches its maximum velocity ($\sim 37$ km/s) only at $r \sim 5$-$6$
kpc, a radius larger than the size of the faintest dwarfs. The
unexpectedly large number of galaxies with velocity widths below the
expected characteristic velocity might then just reflect the fact that physically
small galaxies trace the rising part of the halo circular velocity
curve.

This hypothesis can be checked explicitly if spatially-resolved
rotation curves are available, especially for galaxies where the
inclination is well constrained by good photometry or by
integral-field velocity data. Unfortunately, dwarf galaxies are
typically unresolved in single-dish cm wavelength surveys such as {\sc
  alfalfa}, and the photometric data available are insufficient to
estimate accurately the inclination information needed to turn velocity
widths into circular velocity estimates.

We address these issues here by using a compilation of literature data
for galaxies with spatially-resolved rotation curves and good
photometric data. Since our interest lies in the scale of dwarfs, the
dataset concerns mainly relatively isolated dwarf irregular
galaxies drawn from eight recent studies. The data are heterogeneous,
but they cover a wide range of galaxy stellar mass, from roughly
$10^6$ to $10^{10}\, M_\odot$, and should therefore provide insight
into whether halo masses are in agreement with model predictions. This
paper is organized as follows. We describe the data compilation in
Sec.~\ref{SecData}, present our results in Sec.~\ref{SecResults}, and
summarize our main conclusions in Sec.~\ref{SecConc}.

\begin{center} 
\begin{figure*} 
\includegraphics[width=84mm]{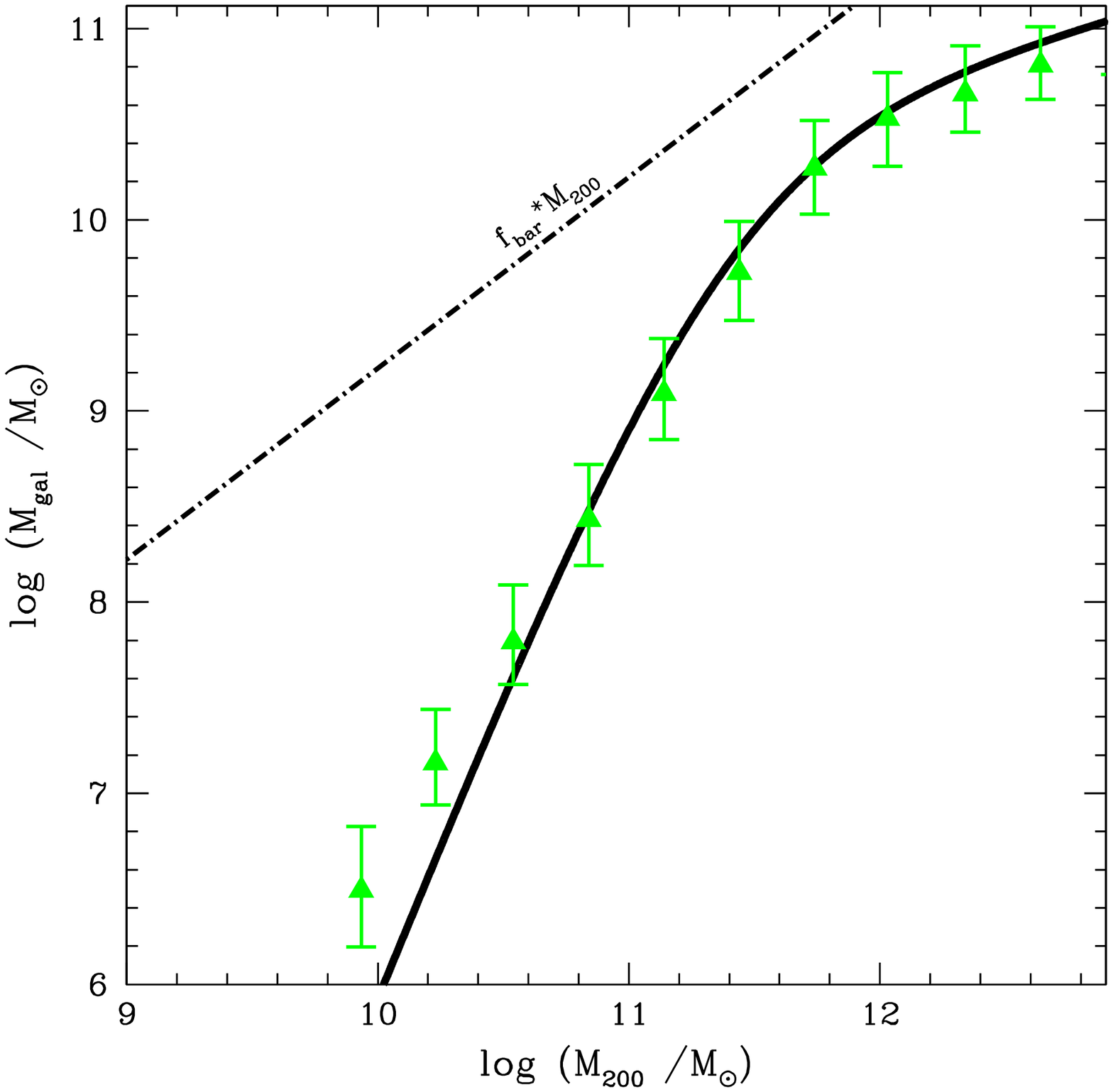} 
\includegraphics[width=84mm]{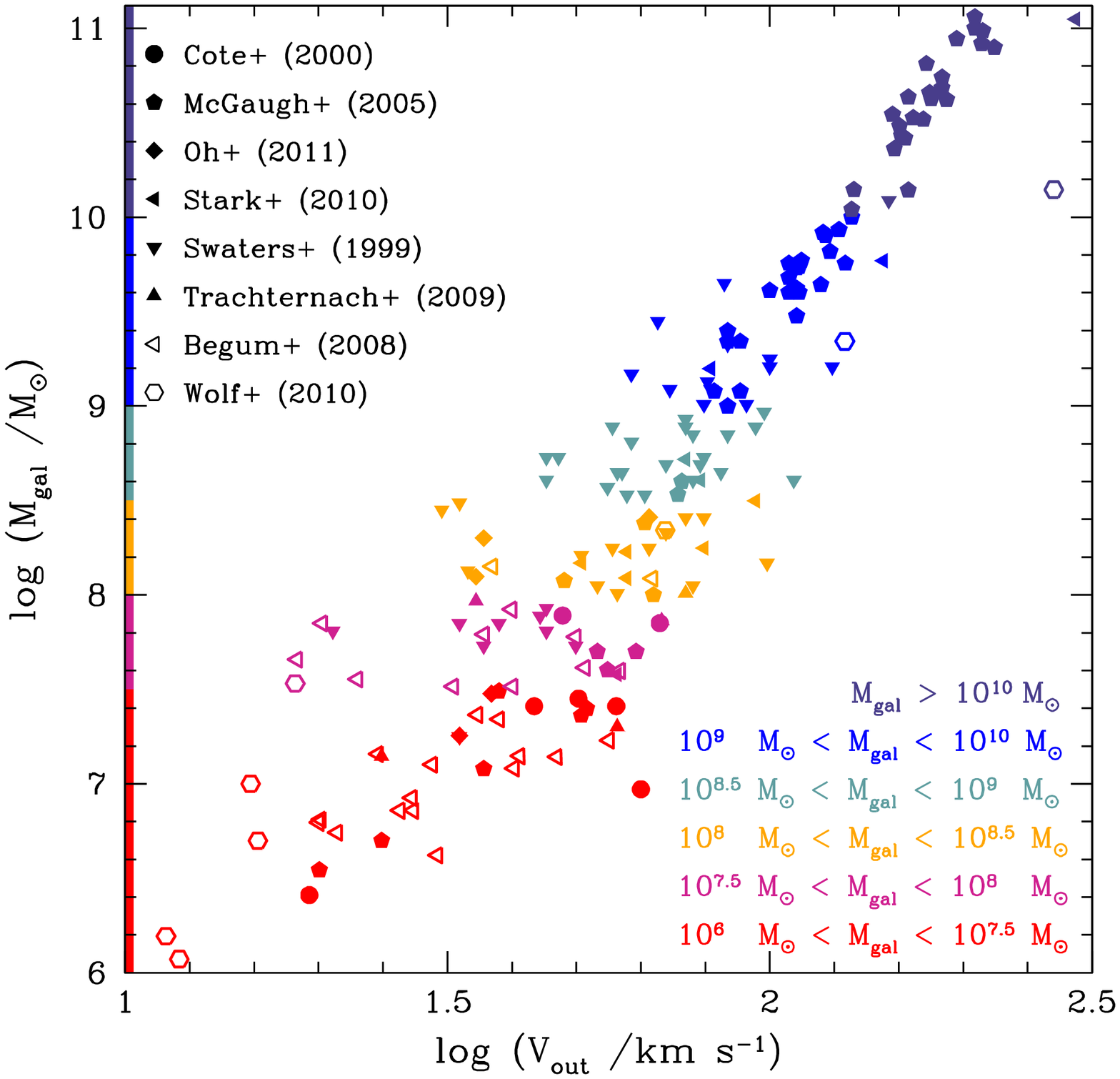} 
\caption{{\it Left:} The halo virial mass vs galaxy stellar mass
  relation derived by \citet{Guo2010} using abundance-matching
  techniques (solid line). Results from the semianalytic model of
  \citet{Guo2011} are shown by the solid triangles. Note how steep the
  relation becomes at the faint end, implying that essentially no
  galaxies with $M_{\rm gal}>10^6 \, M_\odot$ should form in halos
  with mass below $\sim 10^{10} \, M_\odot$. The dot-dashed line
  indicates the baryonic content of a halo according to the latest
  estimates of the universal baryon fraction, $f_{\rm
    bar}=0.171$. {\it Right:} The ``Tully-Fisher'' relation for a
  sample of nearby galaxies. Data are compiled from the sources listed
  in the figure label. Stellar masses are taken from each paper, when
  given, or estimated from their absolute magnitudes and colors as
  described in the text. Rotation velocities correspond to the
  outermost point of the published rotation curve, except for the data
  of \citet{Wolf2010}, which correspond to circular velocities at the
  stellar half-mass radius.  Note that the relation between rotation
  velocity and stellar mass is well approximated by a single power-law
  despite the strongly non-linear $M_{\rm gal}$ vs $M_{200}$ relation
  shown in the left panel.}
\label{FigGuoTF}
\end{figure*}
\end{center}

\section{The Dataset}
\label{SecData}

The main data used in our analysis are HI rotation curves and stellar
masses (or absolute magnitudes) of galaxies compiled from the
literature. The sample we use contains $7$ galaxies from
\citet{Cote2000}, $69$ from \citet{McGaugh2005}, $29$ from
\citet{Begum2008b}\footnote{Complementary information for these
    galaxies was taken from the previous and more extended sample
    presented in \citet{Begum2008a}.}, 
  $5$ from \citet{Oh2011}, $70$ from \citet{Swaters2009}, $5$ from
  \citet{Trachternach2009} and $25$ from \citet{Wolf2010}. We also
  include the $11$ galaxies from \citet{Stark2009} not in the previous
  samples.

We are mainly interested in the total dark mass enclosed by a galaxy,
so in practice we shall use the {\it outermost} point of the rotation
curve, characterized by the radius, $r_{\rm out}$, and rotation
velocity, $V_{\rm out}=V_{\rm rot}(r_{\rm out})$. In most cases, this
is also the maximum rotation velocity measured for the galaxy, since
rotation curves tend to be either rising or flat in the outer
regions. In the rare cases of galaxies with peculiar rotation curves,
such as a steeply-declining outer portion (suggestive of a warp), we
choose instead the radius and velocity of the maximum of the rotation
curve. As we shall see below, this is a conservative choice for the
purpose of our analysis.

The galaxies from \citet{Wolf2010} lack
rotation curve data, but these authors provide the total mass,
$M_{1/2}$, enclosed within the half-light radius, $r_{1/2}$: we shall
adopt then those radii and corresponding velocities,
$V_{1/2}=(GM_{1/2}/r_{1/2})^{1/2}$, as estimates of $r_{\rm out}$, and
$V_{\rm out}$, respectively. The full rotation curves of galaxies
taken from \citet{Begum2008a,Begum2008b} are not yet published, but
the authors list $r_{\rm out}$ and $V_{\rm out}$ in their papers.

Our analysis also makes use of the total stellar mass of the galaxies,
$M_{\rm gal}$, for which we adopt the values quoted in the papers from
which the data are taken. When these are not available, we estimate
stellar masses from the $B$, $R$ or $V$-band absolute magnitude,
assuming \citet{Bell2001} mass-to-light ratios consistent with the
average colors of galaxies in our sample: $\gamma_B=0.5$, $\gamma_R=1$
and $\gamma_V=2$ in solar units. We emphasize that this is not
critical for our analysis, since most dwarfs are heavily dark matter
dominated: we have experimented with increasing and decreasing
$\gamma$ by a factor of two and none of our conclusions are affected
by such changes.

\section{Analysis}
\label{SecResults}

The solid curve in the left panel of Fig.~\ref{FigGuoTF} shows the
galaxy-halo mass relation derived by \citet{Guo2010} assuming that the
abundance of dark halos ranked by virial mass, $M_{200}$, can be
matched monotonically to the abundance of galaxies ranked by stellar
mass, $M_{\rm gal}$. Despite the simplicity of this abundance-matching
technique, more sophisticated semianalytic modeling \citep{Guo2011}
actually yields very similar results, as shown by the solid triangles in
the same figure. A dot-dashed curve indicates the galaxy mass
corresponding to all available baryons within the virial radius,
assuming the universal baryon fraction, $f_{\rm
  bar}=\Omega_b/\Omega_M=0.171$.

The $M_{\rm gal}$-$M_{200}$ relation in Fig.~\ref{FigGuoTF} shows
clearly the sharp decline in galaxy formation efficiency with
decreasing halo mass alluded to in Sec.\ref{SecIntro}: the baryonic
mass of a $M_{200}=10^{11} \, M_\odot$ halo is $f_{\rm bar} \,
M_{200}=1.7 \times 10^{10} \, M_\odot$ but it typically hosts a $10^9 \,
M_\odot$ galaxy containing $\sim 6\%$ of its baryons in the form of
stars. On the other hand, a $10^6 \, M_\odot$ galaxy inhabiting a
$10^{10} \, M_\odot$ halo would contain just $0.06\%$ of its available
baryons.

\begin{center} 
\begin{figure*} 
\includegraphics[width=84mm]{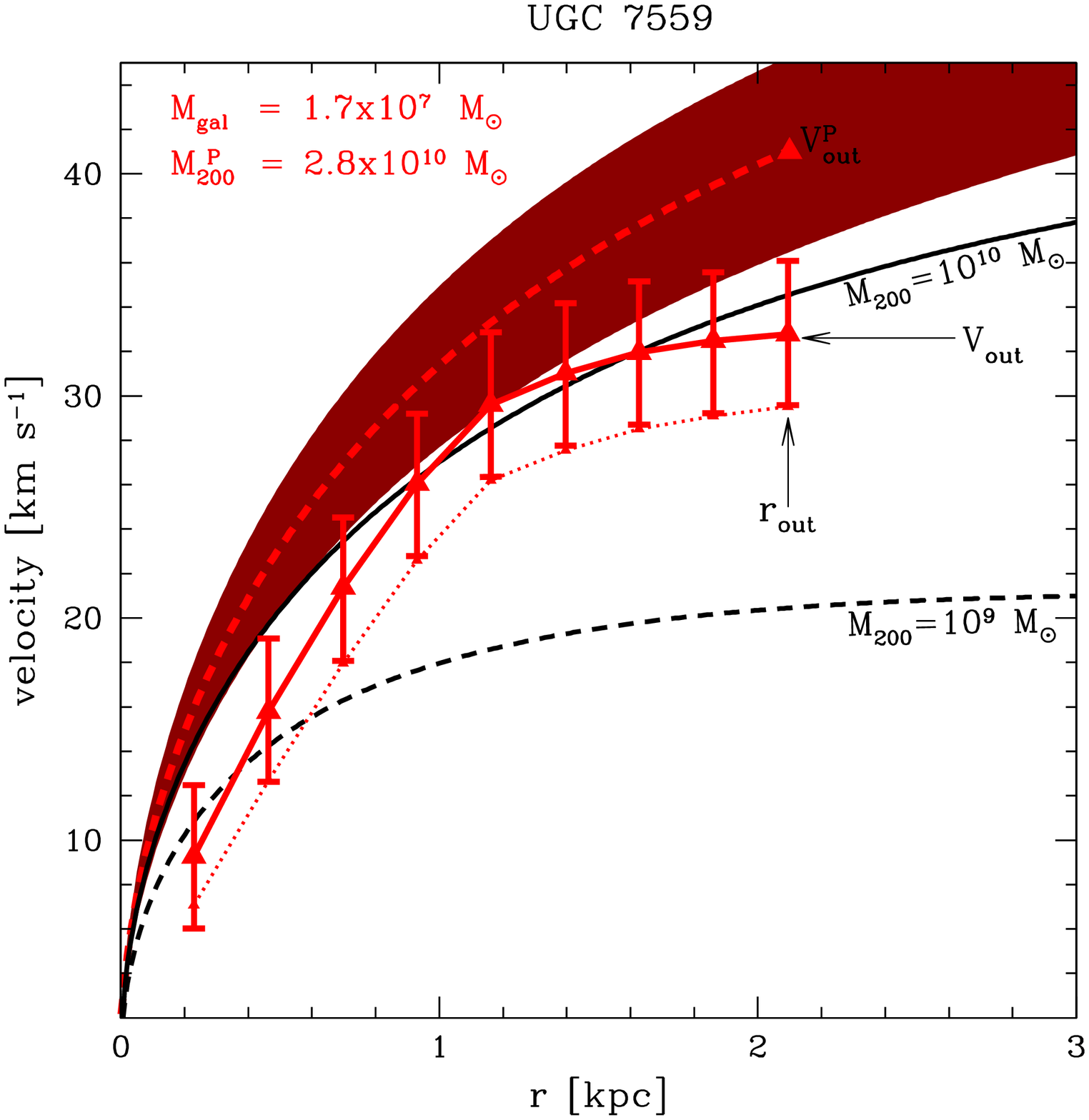} 
\includegraphics[width=84mm]{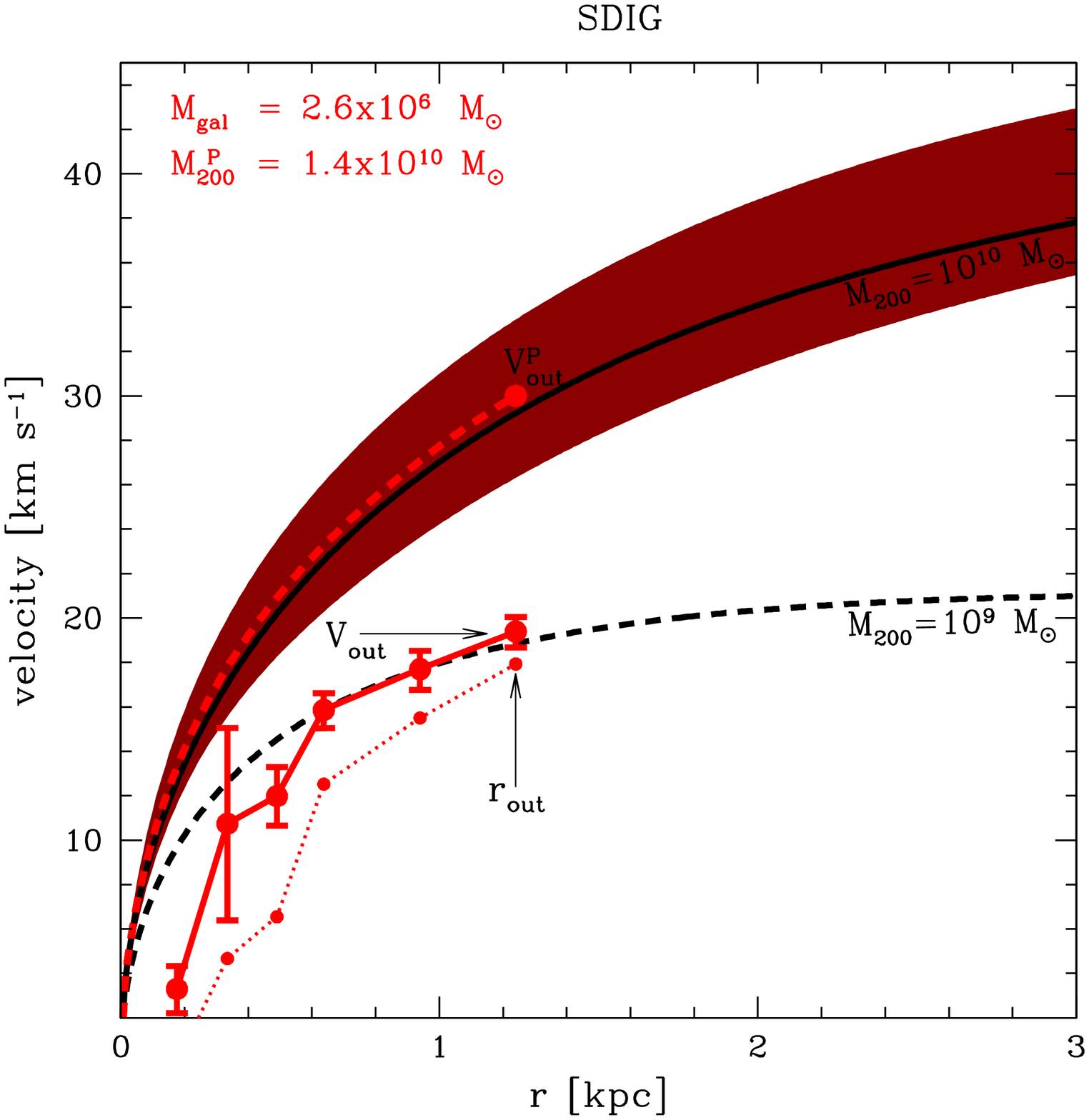} 
\caption{The rotation curve of two dwarf galaxies in our sample: UGC
  7559 \citep{Swaters1999} and the Sagittarius Dwarf Irregular
  \citep[SDIG,][]{Cote2000}.  Filled symbols with error bars reproduce
  the published rotation curve; a dotted line indicates the
  contribution of the dark halo, which dominates the enclosed mass at
  the outermost measured point, $r_{\rm out}$. The black lines
  indicate the circular velocity profile of a $10^{10}$ and a $10^9 \,
  M_\odot$ NFW halo of average concentration, $c=10.8$ and $c=13.4$,
  respectively. The red dashed curve in each panel shows the circular
  velocity profile expected if the NFW halo mass is chosen to match
  the $M_{\rm gal}$ vs $M_{200}$ relation of \citet{Guo2010} given in
  the left panel of Fig.~\ref{FigGuoTF}. The shaded area shows the
  result of varying by $\pm 1\sigma$ the concentration assumed for the
  halo. The abundance-matching model suggests that circular velocities
  should not lie below the black solid line.  Within the uncertainties
  UGC 7559 seems to match this constraint, which is, on the other
  hand, clearly violated by SDIG.}
\label{FigRotCur}
\end{figure*}
\end{center}

\begin{center} 
\begin{figure*} 
\includegraphics[width=0.475\linewidth]{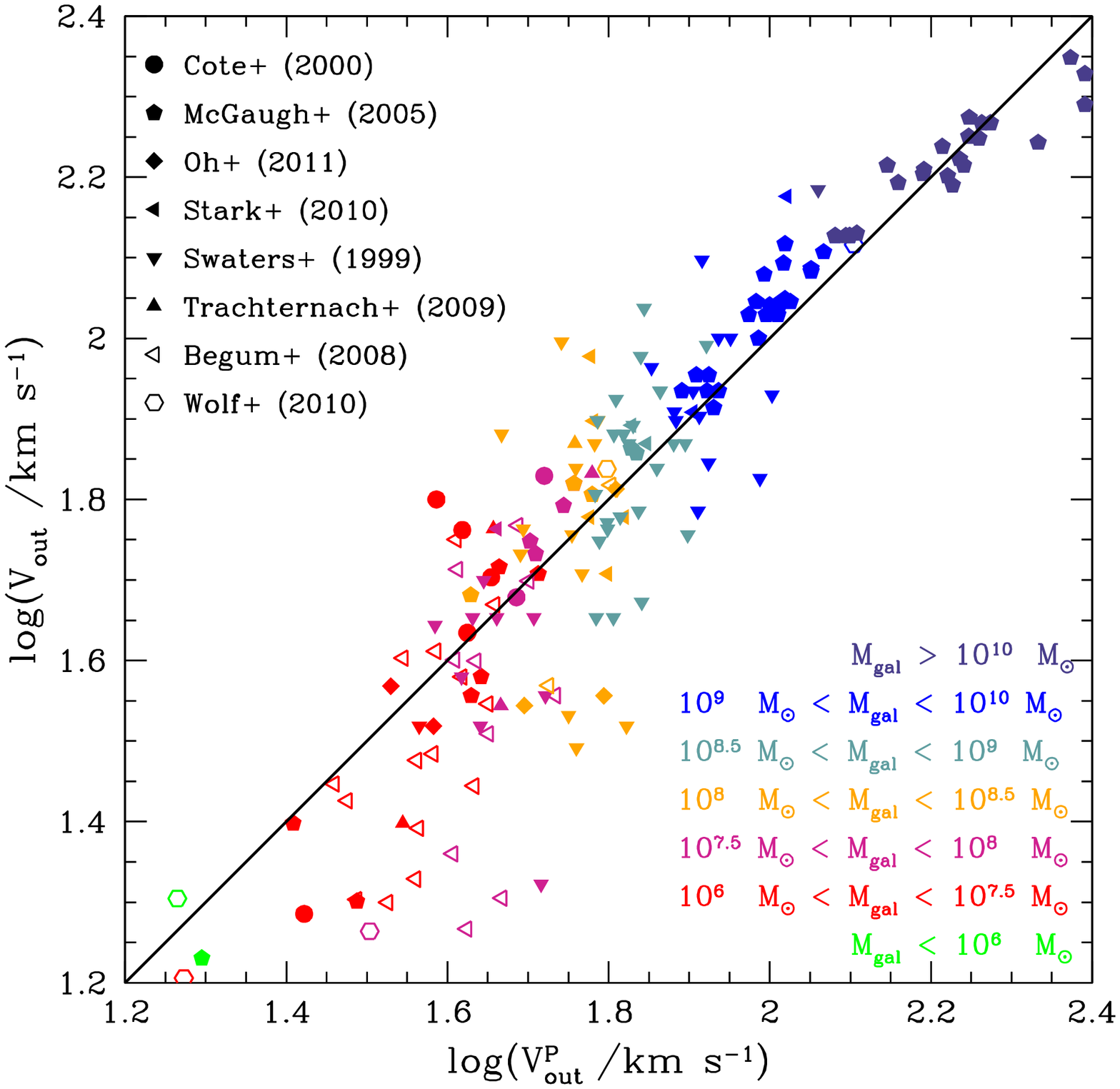} 
\includegraphics[width=0.475\linewidth]{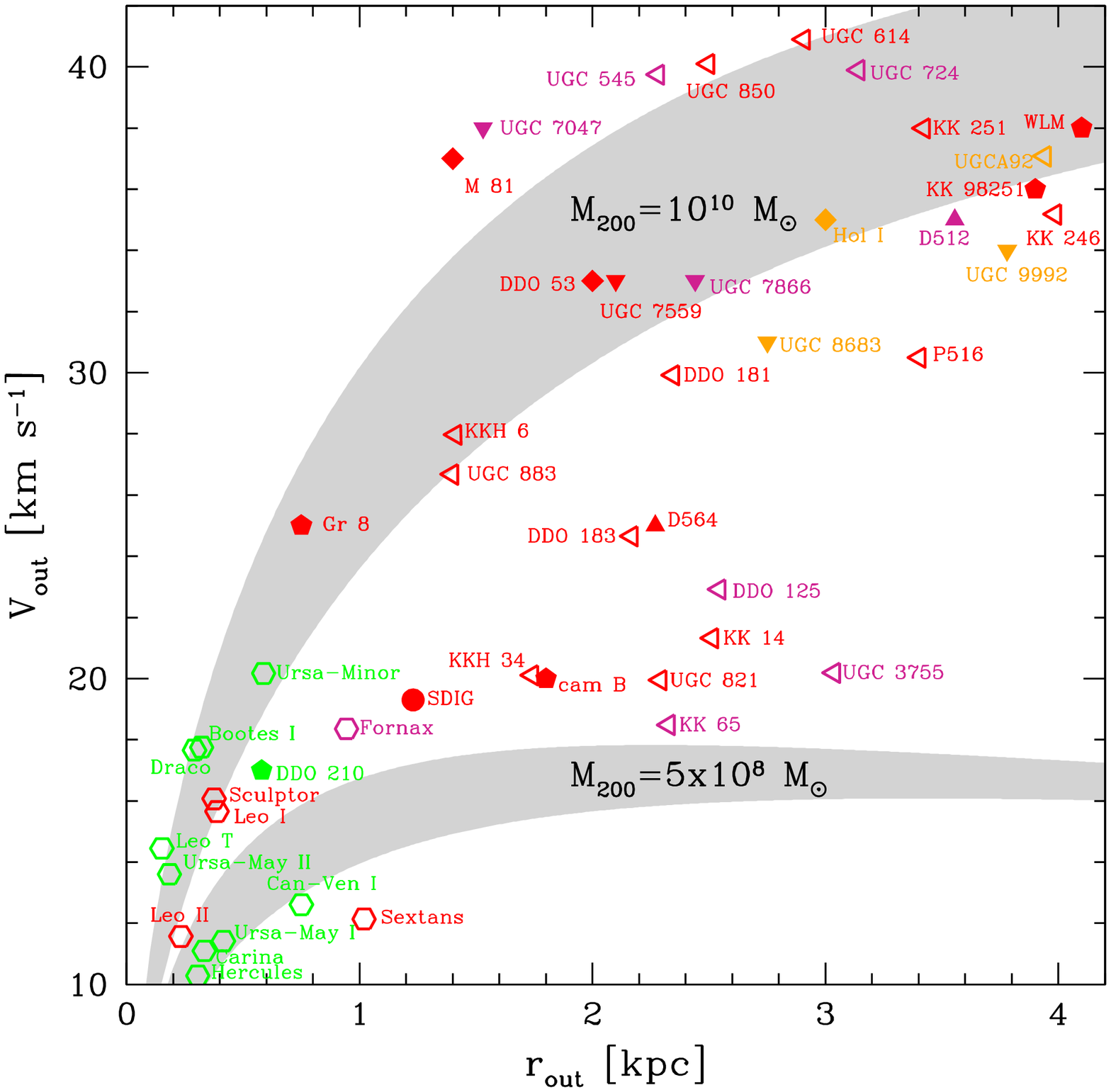} 
\caption{ {\it Left panel:} Outermost rotation velocity, $V_{\rm
    out}=V_{\rm rot}(r_{\rm out})$, measured for each galaxy in our
  sample vs $V_{\rm out}^P$, its predicted value assuming that the
  halo mass is given by the $M_{\rm gal}$ vs $M_{200}$
  abundance-matching relation of Fig.~\ref{FigGuoTF}. Note that the
  faintest dwarfs tend to have velocities well below those expected
  from the model, implying that they inhabit halos less massive than
  expected. {\it Right:} The outermost point of the rotation curve of
  a sample of dwarf galaxies compiled from the
  literature. Abundance-matching arguments suggest that all points
  should lie {\it on or above} the shaded area labeled
  $M_{200}=10^{10}\, M_\odot$. This is clearly not the case. Instead,
  $17$ out of the $44$ galaxies with $V_{\rm outer} < 35 $ km/s
  enclose masses within $r_{\rm out}$ more than a factor of $2$ lower
  than predicted.  The same is true for the faintest dwarfs in our
  sample: roughly $45\%$ of all galaxies with $10^6<M_{\rm gal}<10^7
  \, M_\odot$ have masses that deviate by a similar amount from the
  expected values. If there is a minimum halo mass for dwarf galaxy
  formation, the data imply that it cannot be much higher than $\sim
  5\times 10^8 \, M_\odot$.}
\label{FigRoutVout}
\end{figure*}
\end{center}

Thus, most dwarf galaxies (understood here as those with $M_{\rm gal}
\simlt 10^9 \, M_\odot$) should be surrounded by halos in a fairly
narrow range of mass, spanning less than a decade in $M_{200}$, or
just over a factor of $2$ in circular velocity. Little sign of such
characteristic scale is seen in the Tully-Fisher relation of
galaxies in our sample. This is shown in the right-hand panel of
Fig.~\ref{FigGuoTF}, where we plot the outermost rotation velocity,
$V_{\rm out}$, vs $M_{\rm gal}$ for galaxies in our sample. No obvious
sign of convergence to a characteristic velocity is seen in these
data, which span roughly 5 decades in galaxy mass \citep[see
also][]{McGaugh2010}.

In the $\Lambda$CDM scenario, the nearly self-similar structure of
dark matter halos allows an independent probe of halo mass based on
the rotation curve of a galaxy. We illustrate this in
Fig.~\ref{FigRotCur}, where the thick solid and thick dashed black
curves in each panel indicate the circular velocity profiles of two
NFW halos with virial mass $10^{10}$ and $10^9 \, M_\odot$,
respectively. The concentration parameter of each halo ($c=10.8$ and
$13.4$, respectively) is chosen to be consistent with the results of
\citet{Neto2007}, corrected to the latest values of the cosmological
parameters following \citet{Duffy2008}. The point to note here is that
the more massive a halo the higher its circular velocity is at {\it
  all} radii. The difference is small at small radii (all circular
velocities approach zero at the origin) but it becomes more
appreciable further out. Rotation curves that extend far enough out in
radius are therefore telling probes of the virial mass of the halo.

\begin{center} 
\begin{figure*} 
\includegraphics[width=84mm]{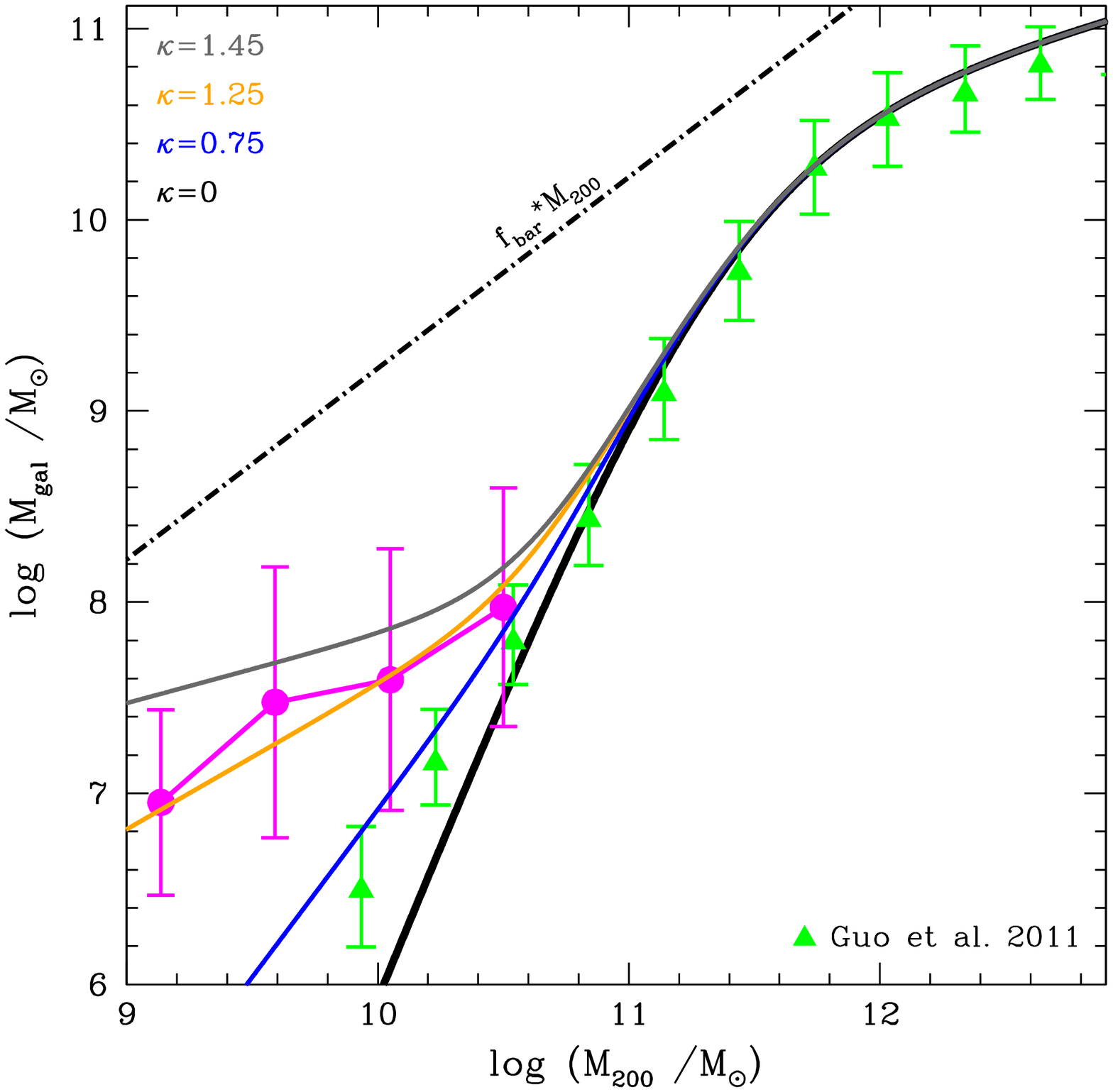} 
\includegraphics[width=84mm]{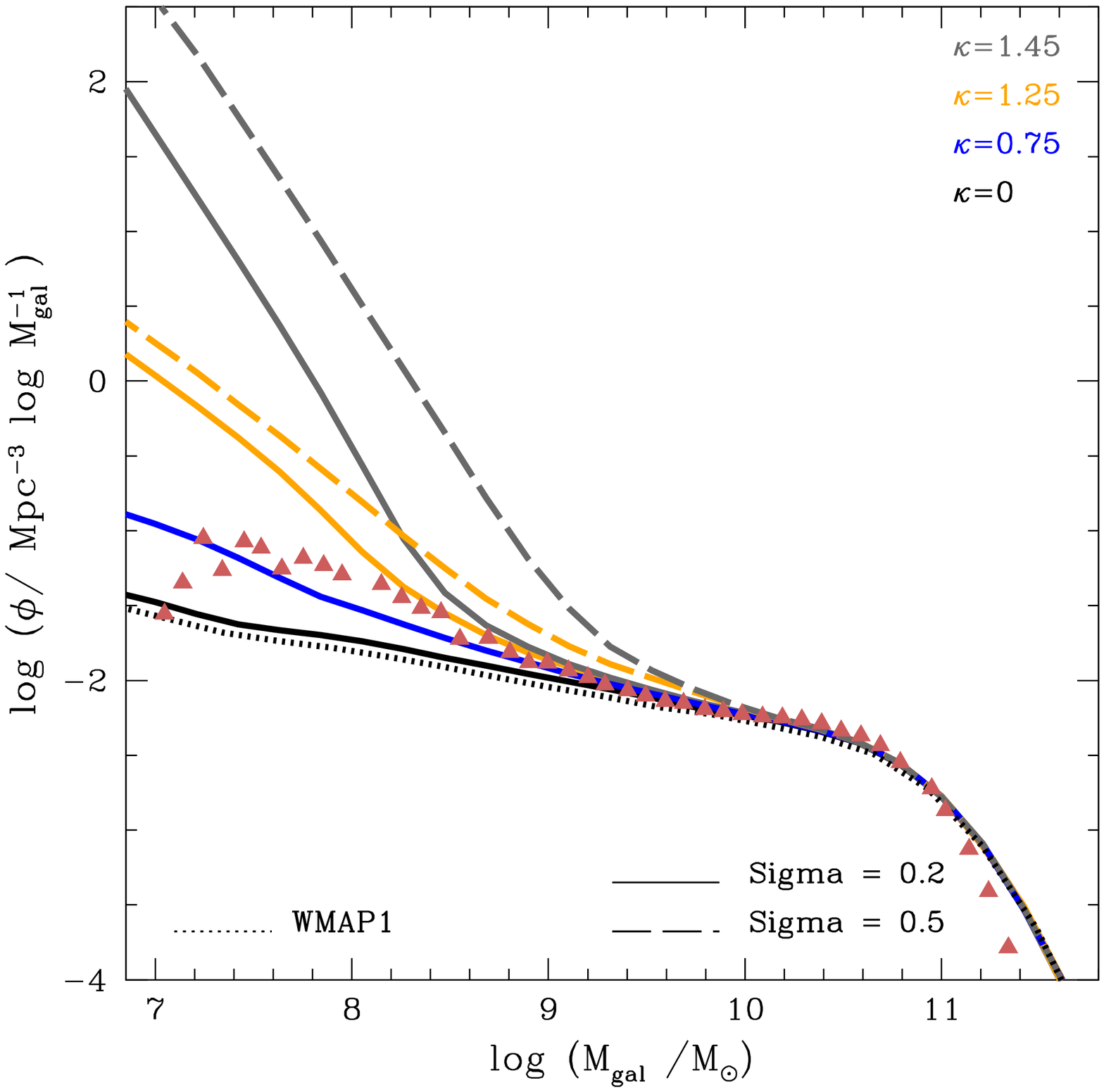} 
\caption{{\it Left:} Galaxy stellar mass-halo virial mass
  relation. The black solid line indicates the abundance-matching
  model of \citet{Guo2010}; solid triangles correspond to the
  semianalytic model of \citet{Guo2011}. The dot-dashed line indicates
  the total baryon mass of a halo according to latest estimates of the
  universal baryon fraction, $f_{\rm bar}=\Omega_{\rm bar}/\Omega_{\rm
    M}$. The magenta curve shows the average galaxy
  mass-halo mass relation derived from 
  dwarf galaxies in our sample. Halo masses of individual galaxies are
  computed by fitting NFW halos to the kinematic data shown in the
  right-hand panel of Fig.~\ref{FigRoutVout}. Circles indicate
  the average in each halo mass bin; the error bar indicates the
  dispersion, computed after $3\sigma$ clipping a few
  outliers. Colored solid curves correspond to various values of the
  parameter $\kappa$ introduced in eq.~\ref{EqMgMh}; $\kappa=0$
  corresponds to the abundance-matching relation, higher values
  correspond to shallower halo mass dependence of galaxy mass. {\it
    Right:}   Galaxy stellar mass function corresponding to the
  various $M_{\rm gal}$-$M_{200}$ relations shown in the left-hand
  panel, contrasted with the observational estimates of
  \citet{Baldry2008} (points with error bars). Solid curves are computed assuming a
  uniform scatter of $0.2$ dex in the galaxy mass-halo mass relation;
  dashed curves assume $0.5$ dex scatter. We assume the halo mass
  function of \citet{Jenkins2001}, corrected to cosmological
  parameters consistent with the latest WMAP measurements.   
  For reference, we also show with a  dotted line the result of adopting  
  cosmological parameters from the   1st-year WMAP  data analysis. Note 
  the large excess of dwarf  galaxies expected if the galaxy-halo mass 
  relation is as shallow as  that suggested by the dwarf kinematic data.}
\label{FigLF}
\end{figure*}
\end{center}

Fig.~\ref{FigRotCur} shows as well the rotation curves of two dwarf
galaxies. The left panel shows UGC 7559, a galaxy with stellar mass
$M_{\rm gal}\sim 1.7\times 10^7 \, M_\odot$ \citep[][]{Swaters1999};
the panel on the right shows the $M_{\rm gal} \sim 2.6\times 10^6 \, M_\odot$
Sagittarius Dwarf Irregular \citep[SDIG,][]{Cote2000}. The published
rotation curves are shown by the solid symbols with error bars; the
smaller symbols connected by a dotted line shows the circular velocity
profile once the contribution of the baryons (gas+stars) has been
discounted. These two galaxies, like most faint galaxies in our
sample, are clearly dark matter dominated in the outer regions. The velocity at the
outermost point of the rotation curve, $V_{\rm out}$, in particular,
depends almost entirely on the enclosed dark mass within $r_{\rm
  out}$, with little contribution from the baryons. 

The red dashed lines in Fig. ~\ref{FigRotCur} indicate the circular
velocity profile\footnote{The prediction assumes an NFW halo
  contracted to account for the effect of the baryonic component
  following \citet{Abadi2010}. This correction is in practice
  negligible for dwarf galaxies since they are almost completely dark
  matter dominated.} expected if the halo mass of each galaxy were to
coincide with the abundance-matching prediction shown in the left
panel of Fig.~\ref{FigGuoTF}. According to this model, the total mass
of the halo inhabited by UGC 7559 ought to be $\sim 2.8\times
10^{10}\, M_\odot$. The upper and lower limits of the shaded region
corresponds to varying the concentration around the average by $\pm
20\%$, the rms scatter at fixed mass reported by \citet{Neto2007} in
their analysis of thousands of halos in the Millennium Simulation. A
similar procedure is followed in the right-hand panel to shade the
region where the rotation curve of SDIG would be expected to lie if
its surrounding halo mass is $\sim 1.4 \times 10^{10} \, M_\odot$, as
suggested by the abundance-matching analysis.

As Fig.~\ref{FigRotCur} makes clear, UGC 7559 is in rough agreement
with the model expectation. Its rotation curve reaches a maximum of
$V_{\rm out} \sim 32$ km/s, at $r_{\rm out}=2.1$ kpc, only slightly
below the $V_{\rm out}^P \sim 40$ km/s expected at that radius
according to the model. The same is not the case for SDIG, whose
paltry peak rotation speed is just $V_{\rm out}\sim 19$ km/s, well
below the $V_{\rm out}^P \sim 30$ km/s expected at
the outermost radius, $r_{\rm out}=1.3$ kpc. 

This is clear indication that the SDIG halo mass is well below the
abundance-matching expectation: a naive fit of the rotation curve
yields $M_{200} \sim 10^{9} \, M_\odot$, a factor of $10$ below the
mass expected from abundance-matching considerations.  Unless the
rotation curve measurements are grossly in error, which we deem
unlikely, it is difficult to evade the conclusion that SDIG truly
inhabits a halo of mass much lower than expected from the model.  Note
that having a spatially-resolved rotation curve that probes a large
radial range is crucial to this conclusion. For example, if the data
available were just a rotation velocity of $19$ km/s from unresolved
data, or if that velocity was reached within, say, $500$ pc, it would
be difficult to discount the possibility that SDIG might inhabit a
much more massive halo.

Could SDIG be instead surrounded by a halo of unusually low
concentration? Indeed, a $M_{200}=10^{10} \, M_\odot$ halo with $c=5$
($3\sigma$ below the average) would match the observed ($r_{\rm
  out}$,$V_{\rm out}$) for this galaxy. If this were true, it would
mean that SDIG is a rare outlier, a possibility that may be checked
by considering the remainder galaxies in our sample.

The results are displayed in Fig.~\ref{FigRoutVout}, where we show, in
the left panel, the measured outermost velocities versus the
velocities predicted (at each value of $r_{\rm out}$) assuming halo
masses derived from the abundance-matching $M_{\rm gal}$ vs $M_{200}$
relation. Although massive galaxies seem to be in good agreement with
the model, those with stellar masses below $\sim 3\times 10^{7} \,
M_\odot$ (and also a few more massive ones) have velocities that fall
systematically below the expected $\sim 30$ km/s corresponding to a
halo mass of $\sim 10^{10} M_\odot$. 

About $17\%$ of galaxies in our sample with $10^7<M_{\rm
  gal}/M_\odot<10^8$ have enclosed masses (within $r_{\rm out}$) more
than a factor of $2$ smaller than expected from the
abundance-matching model. This fraction increases to $45\%$ when
considering galaxies with $10^6 < M_{\rm gal}/ \, M_\odot<10^7$,
ruling out the possibility that galaxies like SDIG are just rare
exceptions.

The right-hand panel of Fig. ~\ref{FigRoutVout} illustrates the
problem in a slightly different way. Here we show the outermost point
of the rotation curves ($r_{\rm out}$,$V_{\rm out}$) of galaxies in
our sample and compare them with the rotation curves expected for NFW
halos of virial mass $10^{10}\, M_\odot$ and $5\times 10^8 \,
M_\odot$, respectively. (Shaded regions correspond to varying the
concentration by $\pm 20\%$, as in Fig.~\ref{FigRotCur}.) There are
clearly many dwarf galaxies, like SDIG, with rotation curves that fall
well below the boundaries imposed by the circular velocity of a halo
as massive as $10^{10}\, M_\odot$.

What could be going on? One possibility is that the interpretation of
the data is incorrect. The rotation velocity of neutral gas in dwarf
irregulars is not a direct measure of the circular velocity, and must
be corrected for the partial support provided by gas pressure, by the
presence of non-circular motions, and by the non-negligible velocity
dispersion of the gas. These corrections are uncertain, and although
they are attempted in most published studies, they may require
revision when better data and more sophisticated modeling are
available. Indeed, the data available in the literature on dwarf
irregulars are highly heterogeneous and of varying quality. For
example, many of the galaxies in our sample taken from
\citet{Begum2008a,Begum2008b} have no published rotation curves (our
analysis uses only their tabulated values of $V_{\rm out}$ and $r_{\rm
  out}$), so it is difficult to assess their reliability.  We have
labeled individually each galaxy in the right panel of
Fig.~\ref{FigRoutVout} in an attempt to encourage further
observational scrutiny of the systems responsible for the challenge we
highlight here.

A second possibility is that baryonic effects such as supernova-driven
gas blowouts \citep[e.g.,][]{Navarro1996}, or gravitational
fluctuations created by star-forming regions \citep[see, e.g.,][and
references therein]{Pontzen2012} might reduce the dark matter content
of dwarf galaxies and alleviate the problem. It is unclear, however,
how baryons in galaxies with stellar masses as small as those of
globular clusters could affect the central regions of a $10^{10} \,
M_\odot$ halo. Although this possibility should be explored more
thoroughly, the outlook does not seem promising \citep[see,
e.g.,][]{Boylan-Kolchin2012,Governato2012}.

Finally, the simplest interpretation is that many dwarf galaxies
inhabit halos of much lower mass than posited by abundance-matching or
semianalytic models.  This,  however, is inconsistent with a
  shallow faint end of the galaxy stellar mass function, as shown in
  Fig.~\ref{FigLF}. The magenta circles in the left-hand panel of
this figure show the average galaxy-halo mass relation derived by
assigning to each dwarf galaxy in our sample a halo mass, $M_{200}$,
consistent with its value of $r_{\rm out}$ and $V_{\rm out}$.

We estimate $M_{200}$ for all $10^6<M_{\rm gal}/M_\odot <10^9$
galaxies in our sample by finding the NFW halo (contracted to account
for the effects of the baryons) whose circular velocity curve passes
through ($r_{\rm out}$,$V_{\rm out}$) after accounting for the
contribution of the gas and stars in the galaxy. We assume that halos
follow the mean mass-concentration relation expected for $\Lambda$CDM
halos (see discussion of Fig.~\ref{FigRotCur} above). Further, we
remove from the analysis the satellites of the Milky Way and M31 since
their mass profiles may have been affected by tides.

The magenta circles in the left-hand panel of Fig.~\ref{FigLF} show
the resulting average galaxy mass as a function of halo mass, together
with error bars denoting the dispersion in each bin computed after
$3\sigma$ clipping a few outliers. The $M_{\rm halo}$ dependence of
galaxy mass is clearly shallower than either the
abundance-matching model results of \citet{Guo2010} (extrapolated to
low halo masses, solid line) or the semianalytic model results of
\citet{Guo2011}, shown with filled triangles. We can parameterize this
deviation by introducing a correction to the functional dependence
advocated by \citet{Guo2010}; i.e.,
\begin{equation}
{M_{\rm gal} \over M_{\rm halo}}= C\,  \biggl[1+\biggl({M_{\rm halo}
  \over M_1}\biggr)^{-2}\biggr]^{\kappa}
\,\biggl[\biggl({M_{\rm halo} \over M_0}\biggr)^{-\alpha} +
\biggl({M_{\rm halo} \over M_0}\biggr)^\beta \biggr]^{-\gamma}, \label{EqMgMh}
\end{equation}
with $C=0.129$, $M_0=10^{11.4} M_\odot$, $M_1=10^{10.65} M_\odot$,
$\alpha=0.926$, $\beta=0.261$, and $\gamma=2.440$. The $M_1$ term in
eq.~\ref{EqMgMh} is our only modification to the relation of
\citet{Guo2010}. The larger the exponent $\kappa$ the shallower the
galaxy-halo mass function is in low mass halos; $\kappa=0$ corresponds
to the original abundance-matching relation (solid line in
Fig.~\ref{FigLF}). We show results for four different values of
$\kappa$ in Fig.~\ref{FigLF}: $\kappa=0$, $0.75$, $1.25$, and $1.45$,
respectively.

A shallow $M_{\rm halo}$-dependence of galaxy mass translates into
large numbers of faint galaxies and, consequently, a steep faint end
of the luminosity function. This is shown in the right-hand panel of
Fig.~\ref{FigLF}, which shows the galaxy stellar mass function
corresponding to each of the choices of $\kappa$ and contrasts them
with the data from \citet{Baldry2008}. Extrapolating the
abundance-matching relation to low-mass galaxies (i.e., adopting
$\kappa=0$; thick solid black curves in Fig.~\ref{FigLF}) actually
yields fewer low-mass galaxies than reported by
\citet{Baldry2008}. Increasing $\kappa$ to $0.75$ fits better the
semianalytic model results of \citet{Guo2011} (solid triangles) and
yields a galaxy stellar mass function in better agreement with
observation down to the smallest galaxy mass probed. 

Increasing $\kappa$ further, as required in order to match the
shallower dependence suggested by our analysis (solid circles),
results in a pronounced excess of low-mass galaxies  over the Baldry et
al data. For $\kappa=1.25$ the model predicts almost one order of
magnitude more $M_{\rm gal}=10^7\, M_\odot$ galaxies than observed, an
excess that increases to more than two decades for
$\kappa=1.45$. These numbers assume that the $M_{\rm gal}$ vs
$M_{200}$ relation has an intrinsic scatter of $0.2$ dex, which is
actually smaller than we find for our dwarf sample (the average
dispersion is $0.6$ dex). An increase in scatter would of course
exacerbate the excess of low-mass galaxies, given the steepness of the
halo mass function. We illustrate this in the right-hand panel of
Fig.~\ref{FigLF}, where the dashed lines assume a scatter of $0.5$
dex. 

 Unless the abundance of dwarfs with $M_{\rm gal}<10^{8.5} \,
  M_\odot$ has been dramatically underestimated in optical surveys (a
  distinct possibility given the difficulty of identifying low-surface
  brightness, faint objects), we are led to the conclusion that it is
not possible to reconcile the low halo masses suggested by the
kinematic data with the faint end of the galaxy stellar mass
function. One solution would be to postulate a mechanism to single out
a small fraction of low mass halos to be galaxy hosts while leaving
dark the vast majority of systems of comparable (or even higher)
mass. The most obvious mechanisms, such as feedback from stellar
evolution and the effects of photoionization, are already included in
the semi-analytic models (see, e.g., Fig.~\ref{FigGuoTF}).  A novel
mechanism  seems required to explain such ``stochasticity'' in the way
dwarf galaxies populate dark matter halos, but has yet to be
identified \citep{Boylan-Kolchin2012}.

\section{Summary}
\label{SecConc}

We have analyzed literature data for a sample of galaxies with
spatially-resolved HI rotation curves and good photometry in order to
place constraints on their halo masses. Our sample spans 5 decades in
galaxy stellar mass, $10^6<M_{\rm gal}/M_\odot< 10^{11}$, with
emphasis on galaxies at the faint end. We focus the analysis on
comparing the (mainly dark) total mass enclosed by dwarf galaxies with
expectations based on galaxy formation models and cosmological N-body
simulations.

Contrary to the general prediction of abundance-matching or
semianalytic models of galaxy formation, we find no evidence that
dwarfs of widely differing stellar mass are surrounded by halos that
span a narrow range in mass.  Further, many of the galaxies in our
sample have enclosed masses much lower than expected from halos as
massive as $10^{10}\, M_\odot$, the characteristic halo mass below
which galaxy formation must become extremely inefficient in order to
reconcile a shallow faint end of the galaxy luminosity function with
the steep dark halo mass function on galactic scales.

If the formation of dwarf galaxies with stellar masses exceeding
$10^6\, M_\odot$ extends to halos with masses as low as a few times
$10^8\, M_\odot$ then this would  lead to a very steep faint end
  of the galaxy stellar mass function unless a mechanism is found to
populate halos with galaxies almost stochastically and with extremely
low efficiency. To our knowledge, no obvious candidate exists for such
mechanism.

The difficulties could be alleviated if the measured rotation curves
underestimate substantially the circular velocity of dwarf
galaxies. The magnitude of the correction needed to bring observed
velocities into agreement with the models appears too large for this
to be a viable alternative. Resorting to baryonic processes to reduce
the dark mass enclosed by dwarfs is similarly unappealing, especially
considering that the discrepancy is clearest in the least massive
systems, some of which contain as few stars as a massive globular
cluster. There are simply too few baryons to drive the transfer of
energy needed to push substantial amounts of matter out of the center
of a massive halo. 

  A more prosaic alternative is that current observations have
  missed a large number of faint galaxies, and that the galaxy stellar
  mass function does indeed have a sharp upturn on mass scales below
  $10^{8.5} \, M_\odot$. Should future observations fail to uphold
  this, however, our finding that many dwarf galaxies inhabit halos with virial
  masses well below $10^{10}\, M_\odot$ would add to the list of
  concerns brought about by the surprisingly low halo masses inferred
  for the dwarf spheroidal companions of the Milky Way
  \citep{Boylan-Kolchin2011,Boylan-Kolchin2012,Parry2011} and by the
  unexpectedly shallow velocity-width function found in blind HI
  surveys \citep{Zwaan2010,Papastergis2011}.

A radical view would take the puzzle we note here as indicative
of the need to revise some of the basic tenets of the $\Lambda$CDM
scenario. Models where low mass halos are substantially less
concentrated or less abundant, such as in a universe dominated by warm
dark matter, for example, might help to resolve the
discrepancy. Alternatively, we must concede that our understanding of
how dwarf galaxies form in $\Lambda$CDM halos is primitive at best,
and perhaps flawed. Neither alternative at this point seems
particularly palatable.

\section*{Acknowledgments}
We acknowledge useful discussions with Mike Boylan-Kolchin and Simon White.
IF and LVS are grateful for financial support from the {\it CosmoComp/Marie Curie} network.

\bibliographystyle{mnras}
\bibliography{master}

\end{document}